\documentclass{SCGE}

\begin{document}

\begin{picture}(0,0){\rm
\put(0,-39){\makebox[160truemm][l]{\bf {\sanhao\raisebox{2pt}{.}}
Article {\sanhao\raisebox{1.5pt}{.}}}}}
\put(0,-52){\jiuwuhao {\textcolor[rgb]{0.5,0.5,0.5}{\sf %Nuclear Magnetic Moments
}}}  %%(11月注释：调\textcolor[rgb]{x,x,x}中的数字x越大越灰)
\end{picture}

\def\bm{\boldsymbol}

\def\dl{\displaystyle}
\def\du{\end{document}}
\def\pi{{\uppi}}

\Year{201X} %
\Month{XX} %
\Vol{XX} %  卷号
\No{XX} %  期号
\BeginPage{1} % 起页码
\EndPage{XX} %  止页码
\AuthorMark{{\rm Wang PeiLong}, et al.}
\DOI{XX} % The author doesn't need fill in it.

% \title[short text for running head]{full title}{comments for title}
\title{Study on the temperature dependence of BGO light yield}%标题$^\dagger$

\author[*]{WANG PeiLong}{}%换手动
\author[*]{ZHANG YunLong}{}
\author[]{XU ZiZong}{}
\author[]{WANG XiaoLian}{}

\address[{\rm}]{State Key Laboratory of Particle Detection and Electronics, University of Science and Technology of China, Hefei 230026, China}

\maketitle \vspace{-3.5mm}{\footnotesize\begin{center} Received September 27, 2013; accepted XX XX, 201X%收稿日期
\end{center}}\vspace*{-5mm}

%     Abstract is required.
\begin{center}
\rule{16.5cm}{0.4pt}
\parbox{16.5cm}
{\begin{abstract} The temperature dependence of BGO coupled with photomultiplier tube R5610A-01 was studied in the range of -30$^{\circ}$C$\sim$30$^{\circ}$C. The temperature coefficient of the BGO and R5610A as a whole was tested to be -1.82\%/$^{\circ}$C. And the temperature coefficient of the gain of the R5610A is -0.44\%/$^{\circ}$C which was tested in the same situation using a blue LED. Thus the temperature coefficient of BGO's light yield can be evaluated as -1.38\%/$^{\circ}$C.%摘要
\end{abstract}}
\end{center}\vspace*{-0.6cm}

\begin{center}
\parbox{16.5cm}
{\bf\jiuhao BGO, light yield, temperature dependence, PMT}%关键词
\end{center}

\begin{center}
{}%分类号
\vspace*{-0.1cm}
%\CITA    %%(11月注释：Citation内容自动生成)
\Cit{Wang P L, Zhang Y L, Xu Z Z et al. Study on the temperature dependence of BGO light yield. Sci China-Phys Mech Astron, 201X, XX: 1--XX, doi: XX}%%(11月注释：Citation 内容需手动填写)
\end{center}

%%%%%%%%%%%%%%%%%%%%%%%%%%%%%%%%%%%%%%%%%%%%%%%%%%%%%%%%%%%%
\wuhao\vspace*{1.5mm}

\begin{multicols}{2}

%%%%%%%%%%%%%%%%%%%%%%%%%%%%%%%%%%%%%%%%%%%%%%%%%%%%%%%%%%%%
%% Text of article.
%%%%%%%%%%%%%%%%%%%%%%%%%%%%%%%%%%%%%%%%%%%%%%%%%%%%%%%%%%%%
%    Section headings
\renewcommand{\baselinestretch}{1.08} \baselineskip 12.2pt\parindent=10.8pt

\renewcommand{\thefootnote}

%\section{Introduction}\label{sec:intro}%“sec:intro”是给该节的命名（可换其他），引用时用\ref{sec:intro}
%\sec{1\quad Introduction}

\no %正文

\sec{1\quad Introduction}

%%%%%%%%%%%%%%%%微调
\setlength{\parskip}{-0.1em}
\no Dark Matter Particle Explorer (DAMPE) is a program which focuses on high energy electron and gamma ray detection in space [1,2]. For measuring the particle energy as precisely as possible, the electromagnetic calorimeter is designed to use 14 layers of bismuth germanium oxide (BGO), 308 BGO bars in total, which will have a 31 radiation length in the Z direction. The dimension of the bars is 25$\times$25$\times$600mm. The direction of the BGO bars of each layer is perpendicular to its neighbors. The calorimeter can contain a particle shower of 10TeV at most. Since the whole detector will run in the temperature variation of -25$^{\circ}$C$\sim$25$^{\circ}$C in space, it is very important to study the temperature dependence of the BGO's light yield to provide the basis for data analysis [3,4].

\sec{2\quad Experimental Setup}

\no The programmable temp.$\&$humi. chamber used in the experiment has a capacity of 1000$\times$1000$\times$1000mm with the temperature range from -70$^{\circ}$C to 150$^{\circ}$C. The cooling down and warming up speed of the chamber is 1$^{\circ}$C/min and 3$^{\circ}$C/min respectively. The stability of the temperature in the chamber

\vspace*{1mm}
\noindent\rule{2.5cm}{0.4pt}\\[0.1mm]{\qihao *Corresponding author (email:
wplong@mail.ustc.edu.cn, ylzhang1@mail.ustc.edu.cn)\vspace*{-1mm}\\
}%手动E-mail 地址$\dagger$Recommended by ZHOU JiLin (Editorial Board Member)

\no is $\pm$0.5$^{\circ}$C. The BGO bar used in this experiment is 25$\times$25$\times$300mm, wholly wrapped by a Teflon tape except one end coupled with a PMT. The PMT is coupled with BGO by air gap and is then slightly tightened with the BGO bar by a heat shrink tube to protect it from light. A thermal resistor is inserted between the Teflon tape and BGO's surface. The PMT used in the experiment is a Hamamatsu R5610A-01 which is a circus cage type with a bialkali photocathode. The working voltage of the PMT is 690V supplied by ORTEC 478, and the signals from its anode are taken via ORTEC 142AH preamplifier, ORTEC 671 amplifier and a multi-channels analyzer. The experimental setup and block diagram are shown in Figure 1.

\begin{figure}[H]%multicols环境下不能浮动，会导致图形或者表格丢失。
%只能当前位置（[]中参数必须是大写H），因此需要手动调位置
\centering
\includegraphics[width=8cm]{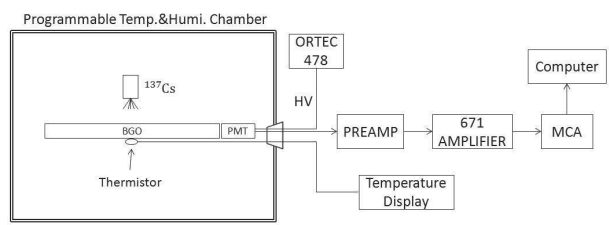}\vspace{-2mm}
\caption{The experimental setup and block diagram for testing the temperature dependence of the BGO's light yield.} %图题
\label{fig:experimental_setup_2}%{}中“fig:example2”为图名，引用时用\ref{fig:example2}
\end{figure}

A radioactive source of ${}^{137}$Cs was located in the middle of the BGO bar where the thermal sensor sat. The spectrum of 662keV was the reference for the BGO's light yield variation. A blue LED (wavelength 476$\sim$495nm) injected light pulses into the PMT's photocathode through an optical fiber to calibrate the temperature dependence of the PMT itself. All of the cables and the optical fiber were connected from a test hole on the chamber.

\sec{2\quad Estimation of the temperature response time from BGO bar's surface to its center}

\no Since it is impossible to insert a temperature sensor into the center of the BGO bar where the scintillation light is produced, the heat equation is used to simulate the heat transfer process to find the delay time that the temperature around the fluorescence point approaches one of BGO's surface recorded by the sensor. To estimate the heat transfer time between the BGO's surface and its center, we suppose that the BGO bar with an equilibrium temperature of 300K is suddenly inserted into an environment with a temperature of 240K. Defining the BGO's center as the origin we can build a Cartesian coordination system in three dimensions. Thus the heat equation in this situation can be expressed as follows.
\begin{equation*}
        \left \{
            \begin{aligned}
             & \frac{\partial{U(x,y,z,t)}}{\partial t}=a^2\triangle{U(x,y,z,t)} \qquad \quad  \ (-12.5<x<12.5,\\
             & \qquad \qquad \qquad  -12.5<y<12.5, -150<z<150, t>0)\\
             %max(0,q_{v,i}^{k}P_{i,j} - \epsilon ), if \  q_{v,j}^{k}>0  \\
             & U(x,y,z,0)=300   \quad (-12.5<x<12.5, -12.5<y<12.5,\\
             & \qquad \qquad \qquad \qquad \qquad \qquad \qquad \qquad -150<z<150)\\
             %max(0,q_{v,i}^{k}P_{i,j} - \epsilon ), if \  q_{v,j}^{k}=0 \  max_{t}(q_{v,t}^{k}P_{t,j}) \geqslant \beta  \\
             & U(-12.5,y,z,t)=240,  \\
             & U(12.5,y,z,t)=240,\\
             & U(x,-12.5,z,t)=240,\\
             & U(x,12.5,z,t)=240,\\
             & U(x,y,-150,t)=240,\\
             & U(x,y,150,t)=240.\\
           % & 0, otherwise
            \end{aligned}
        \right.
    \end{equation*}

The parameter $a^2$ in the first equation is related to the heat transfer coefficient, specific heat capacity and the BGO's density. Since their values don't vary too much in the temperature range of -30$^{\circ}$C$\sim$30$^{\circ}$C, we will just use their values at 270K. The value of $a^2$ can be estimated as follows.
\begin{equation*}
a^2=\frac{k}{c \rho}=\frac{3.34 W/(m\cdot K)}{\displaystyle \frac{358.8 J/(mol\cdot K)}{1245.74 g/mol} \cdot 7.13 g/{cm}^3}=1.627{mm}^2\cdot s^{-1}
\end{equation*}

We can solve the above parabolic partial differential equation using Mathematica to get a numerical solution [5]. And the temperature variation of the BGO center which is the slowest varying point is graphed in Figure 2.

\begin{figure}[H]%multicols环境下不能浮动，会导致图形或者表格丢失。
%只能当前位置（[]中参数必须是大写H），因此需要手动调位置
\centering
\includegraphics[width=7.5cm]{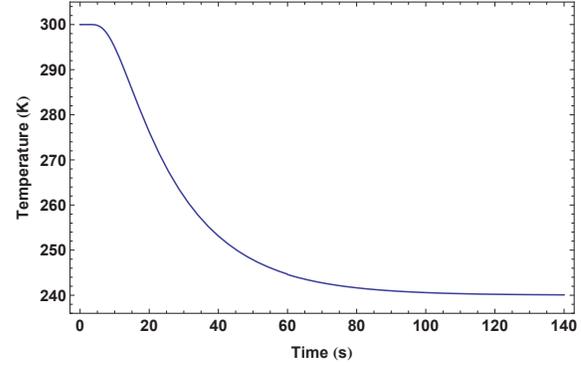}\vspace{-2mm}
\caption{The temperature variation of the BGO center numerically solved by Mathematica.} %图题
\label{fig:one1}%{}中“fig:example2”为图名，引用时用\ref{fig:example2}
\end{figure}

From Figure 2 we can find that it takes about 2 minutes for the BGO's center to vary from 300K to the bar's surface temperature (240K).

\sec{3\quad Data collection and analysis}

\no The test units shown in Figure 1 are ready for data taking at the initial temperature of 22$^{\circ}$C recorded by the thermal sensor, which is the equilibrium temperature of the test chamber and the BGO bar as a whole. Then power the cooling system of the chamber at t=0 to the target temperature of -30$^{\circ}$C. Simultaneously, run the data-taking program at a four-minute repeat period to record the spectrum of 662keV and the instant temperature of the sensor. By fitting the spectrum of each instant temperature the peak channel can be identified [6].

\begin{figure}[H]%multicols环境下不能浮动，会导致图形或者表格丢失。
%只能当前位置（[]中参数必须是大写H），因此需要手动调位置
\centering
\includegraphics[width=8.5cm]{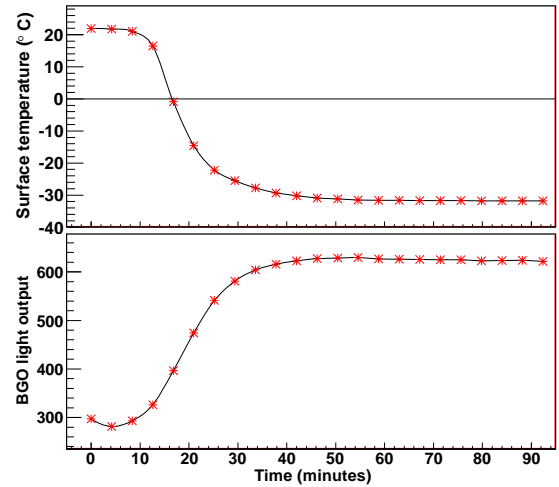}\vspace{-2mm}
\caption{The variation of the BGO's surface temperature and light output with time.} %图题
\label{fig:two_graphs_share_one_Xaxis_4}%{}中“fig:example2”为图名，引用时用\ref{fig:example2}
\end{figure}

Figure 3 shows the instant temperatures of the sensor and its response light yields (peak channels) vary with the cooling down time. It's apparently demonstrated that BGO's light yield roughly doubles as its temperature from 20$^{\circ}$C decreases to -30$^{\circ}$C. The cooling time of the chamber from 20$^{\circ}$C to -30$^{\circ}$C takes about 50 minutes, and the time interval between the temperature's and the spectrum's variation (2$\sim$3 minutes) is in the same order of the heat equilibrium time between the surface and the center of BGO (about 2 minutes, in Figure 2). Therefore Figure 3 could qualitatively describe the temperature dependence of the BGO's light yield.

\sec{4\quad Evaluation of the temperature coefficient of the BGO's light yield}

\no To obtain the correct temperature coefficient of the BGO's light yield, needs to meet the following two conditions: A), the temperature at the surface of the BGO bar recorded by the sensor must be the equilibrium of the domain of the BGO bar where the light emission events happen; and B), the temperature coefficient of the PMT, through which the BGO's light is converted, needs to be tested and subtracted.

For requirement A), the initial temperature was first set to -30$^{\circ}$C, remaining so for 60 minutes so that the BGO bar would reach the equilibrium temperature of -30$^{\circ}$C. The spectrum of 662keV gamma was then taken. The chamber was programmed to the next target temperature of -20$^{\circ}$C, which stayed for another 60 minutes to allow the BGO bar to reach the equilibrium temperature of -20$^{\circ}$C. The spectrum of 662keV gamma was taken. By repeating the data taking in the same manner as the cases of -30$^{\circ}$C and -20$^{\circ}$C, the spectra at the equilibrium temperatures of -10, 0, 10, 20, and 30$^{\circ}$C are taken, too. The fit of the spectra and the peak channels (Figure 4, Top) for each BGO-temperature thus could be obtained. Figure 5 shows that the peak channel's variation with the BGO's temperature is well fitted with a linear function: $Y_{BGO+PMT}$=-8.25($\pm$0.38)T+452.4($\pm$8.3). %The error of the slope is 0.38.

\begin{figure}[H]%multicols环境下不能浮动，会导致图形或者表格丢失。
%只能当前位置（[]中参数必须是大写H），因此需要手动调位置
\centering
\includegraphics[width=8cm]{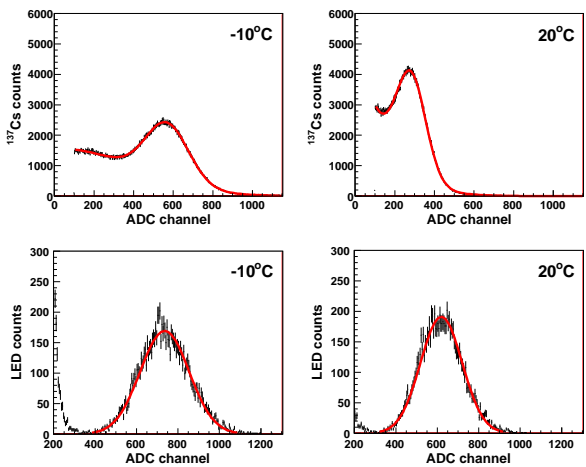}\vspace{-2mm}
\caption{The signals of the BGO and PMT as a whole at the temperatures of -10$^{\circ}$C and 20$^{\circ}$C (Top). The PMT's signal responses to the blue LED at the temperatures of -10$^{\circ}$C and 20$^{\circ}$C (Bottom).} %图题
\label{fig:Cs137_and_LED_four_compare_4}%{}中“fig:example2”为图名，引用时用\ref{fig:example2}
\end{figure}

For requirement B), the PMT was coupled with the optical fiber instead of the BGO and the source of ${}^{137}$Cs was moved out of the chamber. By driving the LED with a constant pulse shape, and repeating the data taking program in the same manner as the BGO's case before, the peak channels of LED spectra (Figure 4, Bottom) for each BGO-temperature (in equilibrium with the PMT) were obtained. As shown in Figure 5, the peak channels of the PMT's gain vary with the temperature which can be well fitted with a linear function: $G_{PMT}$=-3.06($\pm$0.46)T+691.5($\pm$9.7). %The error of the slope is 0.46.

The temperature coefficient is defined as:
\begin{equation*}
C = \frac{slope\ of\ the\ line}{Y(T=0)}.
\end{equation*}
\no Therefore the results are:
\begin{eqnarray*}
C_{BGO+PMT}=-(1.82\pm0.09)\%/^{\circ}C,\\ %\equiv C_{tBP}
C_{PMT}=-(0.44\pm0.07)\%/^{\circ}C. %\equiv C_{tP}
\end{eqnarray*}
\no According to the relation:
\begin{equation*}
Y_{BGO+PMT} (T) = Y_{BGO} (T)\cdot G_{PMT} (T),
\end{equation*}
\no where $Y_{BGO}$ (T) is defined as the light yield of the BGO at a temperature T responding to 662keV gamma, the final equation is
\begin{equation*}
C_{BGO+PMT} = C_{BGO}\cdot C_{PMT}\cdot T^2+C_{BGO}\cdot T+C_{PMT}\cdot T.
\end{equation*}
Since PMT's temperature coefficient $C_{PMT}$ is very small, $T^2$ term could be omitted when the temperature T is close to 0$^{\circ}$C. Thus the final result is
\begin{equation*}
C_{BGO} = C_{BGO+PMT}-C_{PMT}=-(1.38\pm0.11)\%/^{\circ}C.
\end{equation*}

\begin{figure}[H]%multicols环境下不能浮动，会导致图形或者表格丢失。
%只能当前位置（[]中参数必须是大写H），因此需要手动调位置
\centering
\includegraphics[width=8cm]{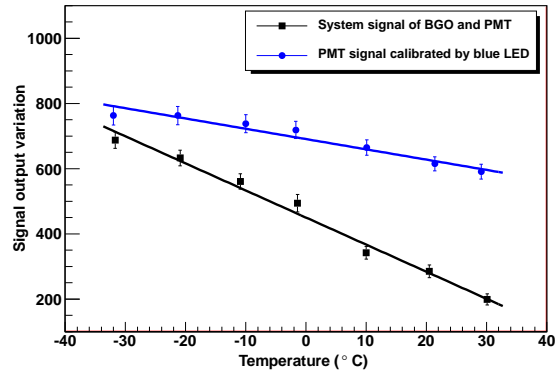}\vspace{-2mm}
\caption{The temperature dependence of the BGO-PMT system and the PMT itself.} %图题
\label{fig:result_BGO_temperature_dependence_7}%{}中“fig:example2”为图名，引用时用\ref{fig:example2}
\end{figure}

\sec{4\quad conclusion}

\no The temperature response of the BGO coupled with the R5610A-01 was studied in the range of -30$^{\circ}$C$\sim$30$^{\circ}$C. The temperature dependence of the gain of the R5610A was tested in the same temperature range using a blue LED. The temperature coefficient of BGO's light yield was evaluated from the data of ${}^{137}$Cs+BGO+PMT combination and LED+PMT only.

\vspace*{2mm} \Acknowledgements{\bahao  We would like to thank Professor Guangshun Huang for some extremely useful discussions. We are also thankful to Mr. Grant Renny for his helpful English assistance. This work is supported by the 973 Program (Grant No. 2010CB833002), and the Strategic Priority Research Program on Space Science of the Chinese Academy of Sciences (Grant No. XDA04040202-4).}

\normalsize \vskip0.3in\parskip=0mm \baselineskip 18pt
\renewcommand{\baselinestretch}{1.1}\footnotesize\parindent=4mm\bahao

%\vskip0.1in \noindent %{\normalsize \bf References}
%\vskip0.1in\parskip=0mm

\REF{1\ }Hooper D, Silk J. Searching for Dark Matter with future cosmic positron experiments. Phys Rev D, 2005, 71: 083503

\REF{2\ }Chang J, Adams Jr J H, Ahn H S, et al. An excess of cosmic ray electrons at energies of 300$-$800 GeV. Nature, 2008, 456: 362--365

\REF{3\ }Melcher C L, Schweitzer J S. Temperature dependence of fluorescence decay time and emission spectrum of bismuth germante. IEEE Trans Nucl Sci, 1985, 32(1): 529--532

\REF{4\ }Isbert J, Adams Jr J H, Ahn H S, et al. Temperature effects in the ATIC BGO calorimeter. Adv Space Res, 2008, 42(3): 437--441

\REF{5\ }Wolfram Mathematica 8. Version 8.0.1.0. Champaign (IL): Wolfram Research. 2011

\REF{6\ }Baccaro S, Barone L M, Borgia B, et al. Precise determination of the light yield of scintillating crystals, Nucl Instr Meth A, 1997, 385(1): 69--73

\end{multicols}

\end{document}